\documentclass[pdflatex,sn-mathphys-num]{sn-jnl}

\usepackage{graphicx}%
\usepackage{multirow}%
\usepackage{amsmath,amssymb,amsfonts}%
\usepackage{amsthm}%
\usepackage{mathrsfs}%
\usepackage[title]{appendix}%
\usepackage{xcolor}%
\usepackage{textcomp}%
\usepackage{manyfoot}%
\usepackage{booktabs}\let\cline\cmidrule%
\usepackage{algorithm}%
\usepackage{algorithmicx}%
\usepackage{algpseudocode}%
\usepackage{listings}%

\theoremstyle{thmstyleone}%
\theoremstyle{thmstyletwo}%

\theoremstyle{thmstylethree}%

\begin{document}

\title[Article Title]{High-Performance Implementation of the Optimized Event Generator for Strong-Field QED Plasma Simulations}


\author[1]{\fnm{Elena} \sur{Panova}}\email{elena.panova@itmm.unn.ru}

\author[1]{\fnm{Valentin} \sur{Volokitin}}\email{volokitin@itmm.unn.ru}

\author[2]{\fnm{Aleksei} \sur{Bashinov}}\email{bashinov@ipfran.ru}

\author[2]{\fnm{Alexander} \sur{Muraviev}}\email{sashamur@gmail.com}

\author[1]{\fnm{Evgeny} \sur{Efimenko}}\email{evgeny.efimenko@gmail.com}

\author*[1]{\fnm{Iosif} \sur{Meyerov}}\email{meerov@vmk.unn.ru}

\affil*[1]{\orgdiv{} \orgname{Lobachevsky State University of Nizhni Novgorod}, \orgaddress{\street{} \city{Nizhni Novgorod}, \postcode{603950}, \state{} \country{Russian Federation}}}

\affil[2]{\orgdiv{} \orgname{Institute of Applied Physics of RAS}, \orgaddress{\street{} \city{Nizhni Novgorod}, \postcode{603950}, \state{} \country{Russian Federation}}}

\abstract{Numerical simulation of strong-field quantum electrodynamics (SFQED) processes is an essential step towards current and future high-intensity laser experiments. The complexity of SFQED phenomena and their stochastic nature make them extremely computationally challenging, requiring the use of supercomputers for realistic simulations. Recently, we have presented a novel approach to numerical simulation of SFQED processes based on an accurate approximation of precomputed rates, which minimizes the number of rate calculations per QED event. The current paper is focused on the high-performance implementation of this method, including vectorization of resource-intensive kernels and improvement of parallel computing efficiency. Using two codes, PICADOR and hi-$\chi$ (the latter being free and publicly available), we demonstrate significant reduction in computation time due to these improvements. We hope that the proposed approach can be applied in other codes for the numerical simulation of SFQED processes.}


\maketitle

\section{Introduction}\label{sec:intro}

Strong field laser-plasma interaction is signified by the initiation of complex and highly nonlinear processes \cite{RevModPhys.78.309} that are not relevant at lower intensities. As a result, numerical modeling has universally become a key tool for SFQED studies. Moreover, the results of numerical modeling often play a crucial role in the design and development of experiments and in the interpretation of experimental results on modern laser facilities \cite{Cole_PRX2018,Poder_PRX2018}. One of the most widespread computational approaches is based on the Particle-In-Cell method (PIC) \cite{birdsall2018plasma}. Within the PIC method plasma is considered as an ensemble of a large number of interacting macroparticles, which represent a number of real physical particles located close within the phase space. The method describes the self-consistent dynamics of charged macroparticles in electromagnetic fields. It is based on mesh discretization of the electromagnetic field followed by integration of the Maxwell equations \cite{taflove2005computational,lehe2018review} and equations of motion \cite{boris1970relativistic,decyk2023analytic}. This setup does not require accounting for any direct interaction between macroparticles, which makes the “macroparticle-mesh” interaction local, which in turn facilitates the parallelization of computations on supercomputers \cite{decyk2023analytic,fonseca2013exploiting,vay2013domain,vshivkov1998parallel} –- a key feature for most state-of-the-art applications \cite{fedeli2022pushing}.

Due to the active development of laser systems \cite{danson.hplse.2019}, and with field intensities approaching the threshold value of $10^{23}$ W/cm$^2$, SFQED processes are becoming more and more relevant as they can not only significantly affect laser-plasma interaction, but qualitatively change its behavior \cite{Piazza_RMP2012,RevModPhys.94.045001,Popruzhenko:2023}. SFQED processes are stochastic and highly nonlinear, which immensely complicates their consideration in simulations. Moreover, SFQED processes can cause the number of particles in a relatively small volume to grow exponentially \cite{bell.prl.2008,fedotov.prl.2010,bulanov2010schwinger,nerush2011laser}, which can lead to the following problems. First and most obvious, this can significantly increase the requirements for memory and computational resources. This is usually combated by resampling (thinning or merging) of the particle ensemble \cite{muraviev.cpc.2021,vranic2015particle,dong2024dynamical}. Second, the distribution of macroparticles in the computational region may become unbalanced, which requires some method of computational load balancing \cite{germaschewski2016plasma,nakashima2009ohhelp,surmin2015dynamic,miller2021dynamic,meyerov2019exploiting}. And third, the conventional SFQED schemes require a significant reduction of the PIC code time step in order to resolve SFQED processes, which leads to great computational tolls. The current paper addresses the latter problem.

Several computational approaches for simulation of SFQED processes have been proposed and implemented over the past decade. However, all of them have an additional theoretical limitation regarding the time step of the simulation. In \cite{volokitin2023optimized} we proposed a fundamentally new approach to modeling of QED cascades, which eliminates the need for redundant time step subdivision and significantly eases the demand for computational resources. The current paper takes the next step in improving performance by presenting a new and optimized open-source implementation of the QED module that effectively utilizes CPU resources through vectorization and parallelization of computations. 

The scientific contribution of this paper is as follows: 

\begin{enumerate}
    \item We present an optimized implementation of the algorithm \cite{volokitin2023optimized}, which speeds up QED simulations by 30-60\%.
    \item We describe and analyze approaches to improving the performance of QED simulations. We hope these methods can be applied in other codes as well. 
    \item We implement and test our methods in two codes: PICADOR and Hi-$\chi$. The latter is publicly available in source codes \cite{hichi}.
\end{enumerate}

\section{Related Works}\label{sec:works}

Studying physics of plasma in ultraintense laser fields implies considering of quantum processes. QED-PIC is one of the most widely used approaches to numerical modeling of QED effects in plasma \cite{ridgers2014modelling,nerush2011laser}, which has been actively developed in recent years. In 2015, we presented a method \cite{gonoskov2015extended} for adopting the SFQED module in PIC codes and implemented it in ELMIS \cite{gonoskov2015extended} and PICADOR \cite{surmin2015dynamic}. The method was based on a modified event generator that models the entire spectrum of incoherent particle emission and employs time-step subcycling in order to resolve QED events. At the same time, an inverse sampling method with the optical depth parameter computing \cite{ridgers2014modelling} became widespread. This approach allows us to avoid sub-stepping but has an additional theoretical limitation on the global time step. These and similar approaches have been implemented in various PIC-codes such as CALDER \cite{lobet_2016}, EPOCH \cite{arber_2015}, PICSAR \cite{fedeli2022picsar}, OSIRIS \cite{grismayer.pre.2017}, PIConGPU \cite{bussmann2013radiative}, Smilei \cite{derouillat2018smilei}, Quill \cite{nerush2011laser}. In 2020, we presented an improved QED module with new polynomial approximations for synchrotron radiation functions \cite{volokitin.jpcs.2020}. In 2021, a comprehensive open-source library PICSAR-QED, optimized for modern heterogeneous cluster systems, was presented and integrated into the PICSAR and WarpX PIC codes \cite{fedeli2022picsar}. In 2022, Guo et al. \cite{guo2022improving} applied a sigmoid approximation to the electron/positron emission probability function. In 2023, Montefiori et al. \cite{montefiori2023sfqedtoolkit} presented an efficient QED-PIC implementation using Chebyshev polynomial approximations. In 2023, we introduced a fundamentally new approach to the modeling of QED cascades \cite{volokitin2023optimized}, which implies the minimum number of possible rate evaluations, either one per event or one per global time step.

\section{Models and Algorithms}\label{sec:model}
 One of the properties of SFQED processes that makes them challenging for numerical implementation is their probabilistic nature.  A particularly fruitful methodology is based on extending the Particle-In-Cell (PIC) method with a Monte-Carlo module that accounts for the SFQED processes via probabilistic generation of macro-photons, electrons and positrons that sample the 3D3P distribution by an ensemble of reasonable size \cite{nerush2011laser, elkina.prstab.2011, gonoskov2015extended, fedeli2022picsar, lobet_2016, arber_2015}. 

Rapid particle production results in an extremely short characteristic temporal interval between SFQED events as compared to the laser wave period and/or other macroscopic scales of the simulated processes. The way to handle computational demands caused by this depends on the method used for sampling SFQED events: \textit{rejection} or \textit{inverse} sampling. 

In the case of \textit{rejection sampling} two random numbers are generated for each time step and for each photon/particle: one that defines the proposed energy of photon/particle to be generated and another that determines whether the proposal is accepted or rejected \cite{nerush2011laser}. This method is particularly simple to implement as it requires the computation of two special functions, more precisely the first and second synchrotron functions~\cite{gonoskov2015extended}. By setting the requirements on the time step and implementing time step sub-cycling in order to meet said requirements, it can be made possible to meet the general rejection sampling requirements for any simulation setting \cite{gonoskov2015extended}. For each particle/photon the described procedure permits the generation of arbitrarily many events (even local cascades) during each global time step, making the whole computational method versatile for studies of extreme processes, e.g. resolving formation and self-consistent dynamics of ultra-dense electron-positron plasma \cite{
efimenko.scirep.2018}. However, the use of rejection sampling has a downside: this method usually requires multiple, and sometimes expensive, computations of the involved special functions \cite{nikishov.spjetp.1964} per accepted SFQED event. 

The alternative approach is \textit{inverse} sampling, in which a distribution is sampled through its cumulative distribution \cite{arber_2015, montefiori2023sfqedtoolkit}. The idea behind the method is the following: for each particle/photon a random number uniformly distributed in the unit interval is generated and the so-called \textit{optical depth} is computed. If we represent the probability of an event \emph{not} happening within the time interval $(t, t + t_1)$ as $P(t, t + t_1) = 1 - \exp\left(-\int_t^{t+ t_1} R(\tau)d\tau\right)$, where $R(\tau)$ is the instantaneous event rate over the particle's trajectory, then the exponent plays the role of a cumulative optical depth. As the particle propagates, the integral in the exponent is computed using first-order Eulerian integration until it reaches the value of $d$, at which point the event is assigned to occur. The energy of the outgoing particles is determined randomly using inverse sampling through pre-computed look-up tables that store the inverse cumulative probability as a two-dimensional function of energy and the quantum nonlinearity parameter $\chi$. A new value of optical depth is then generated in order to determine the time for the next event to happen, and so on. 
In this case the most computationally demanding part is the evaluation of the rate, which is performed only once per particle per global time step. This is done in many state-of-the-art QED-PIC codes, including \cite{arber_2015, fedeli2022picsar}. 

Recently we proposed the implementation of the method based on the inverse sampling methodology with sub-cycling\cite{volokitin2023optimized}. In doing so, within each global time step multiple SFQED events originated from a single particle/photon may occur, including ones produced by secondary particles. After the optical depth is reached, new particles are generated, optical depths are assigned for each of them and the process is continued until the end of the global time step. Then the rate for all particles/photons in the ensemble is re-evaluated, as the global time step often defines the time scale over which the field can no longer be assumed to remain unchanged. For this reason this procedure implies the minimum number of possible rate evaluations: either one per event or one per global time step, depending on whether the optical depth is shorter or longer than the global time step.


The main idea behind the particle-push method, that takes into account QED effects, is presented in Algorithm~\ref{alg:particle_push}. The generation of a QED cascade involves two processes: photon emission from charged particles and particle production by high-energy photons. In the Algorithm~\ref{alg:particle_push}, we only demonstrate a procedure for photons emission. For the production of electron-positron pairs, the concept of the particle-push method is similar, except that the generating photon is eliminated after a pair is created. We assume that the birth time of a particle is an exponentially distributed quantity with an event rate $\lambda$. Therefore, the emission time can be calculated as $\hat t = -\log(r)/\lambda$, where $r$ is a uniformly distributed random number. The Compton and Breit-Wheeler event rates are estimated using their polynomial approximations \cite{volokitin2023optimized}. If no events occur within the time interval $[0, \Delta t]$, where $\Delta t$ is the time step used the particle-in-cell method, the particle is simply advanced to the final time point. Otherwise, the particle is advanced to the calculated time $\hat t$, and a photon or a pair with an appropriate energy is generated. After this, we repeat these steps for the new time interval $[\hat t, \Delta t]$.

\begin{algorithm}[!ht]
\caption{The particle push algorithm for
charged particles with account for QED effects}\label{alg:particle_push}
\lstset{language=pascal,texcl=true,numbers=left,basicstyle=\small\sf,commentstyle=\small\rm,mathescape=true}
\begin{lstlisting}          
procedure ParticlePush($particle$, $generatedPhotons$, $t_{final}$)
    while $particle$.time < $t_{final}$ do
        $\chi$ = Calculate$\chi$($particle$)
        $\lambda$ = CalculateRate($\chi$)
        $\Delta \hat t$ = -log(random())/$\lambda$
        if $particle$.time + $\Delta \hat t$ > $t_{final}$ then
            BorisPusher($particle$, $t_{final}$ - $particle$.time)
            $particle$.time = $t_{final}$
        else
            BorisPusher($particle$, $\Delta \hat t$)
            $particle$.time += $\Delta \hat t$
            $newPhoton$ = PhotonEmission($particle$)
            $generatedPhotons$.Add($newPhoton$)
\end{lstlisting}
\end{algorithm}

In our PIC codes, the particle-push algorithm is applied to each particle in the particle ensemble, including newly generated particles. During the time step $\Delta t$, a charged particle can generate a number of photons, which in turn can generate a number of electron-positron pairs, leading to an electron-positron avalanche. Each generated particle is also pushed forward to the final time point using the procedure described above. The method for processing avalanches is presented in Algorithm~\ref{alg:run_avalanche} and can be used in PIC codes as a particle pusher instead of frequently used Boris pusher. However, it should be noted that such a pusher not only processes existing particles, but also generates new particles, which must be added to the particle ensemble in a proper way.

\begin{algorithm}[!ht]
\caption{The avalanche processing procedure}\label{alg:run_avalanche}
\lstset{language=pascal,texcl=true,numbers=left,basicstyle=\small\sf,commentstyle=\small\rm,mathescape=true}
\begin{lstlisting}
procedure RunAvalanche($particle$, $\Delta t$)
    $t_{final}$ = $particle$.time + $\Delta t$
    $generatedPhotons$ = [], $generatedParticles$ = []
    $countPhotons$ = 0, $countParticles$ = 0
    if $particle$.particleType == PHOTON then
        $generatedPhotons$.Add($particle$)
    else $generatedParticles$.Add($particle$)
    while $countParticles$ != len($generatedParticles$) \
          or $countPhotons$ != len($generatedPhotons$) do
        for $k$ = $countParticles$ to len($generatedParticles$) do
            ParticlePush($generatedParticles$[$k$], $generatedPhotons$, $t_{final}$)
        $countParticles$ = len($generatedParticles$) 
        for $k$ = $countPhotons$ to len($generatedPhotons$) do
            PhotonPush($generatedPhotons$[$k$], $generatedParticles$, $t_{final}$)
        $countPhotons$ = len($generatedPhotons$)
    SaveGeneratedParticles($generatedPhotons$, $generatedParticles$)
\end{lstlisting}
\end{algorithm}

\section{Optimization}\label{sec:optimization}

\subsection{Data structures}\label{subsec:data_structures}

An important aspect of developing high-performance PIC codes is choosing appropriate data structures \cite{barsamian2018efficient}. In this section we discuss the memory representation of an ensemble of particles in QED-PIC codes. Specifically, we consider an effective approach to store and process particles generated by the QED particle pusher during a QED cascade.

There are two main approaches to organize the ensemble of particles in PIC codes. The first and the simpler to implement method involves storing all particles in a single array. However, this method may lead to frequent data conflicts in multi-threaded mode during the current deposition stage, when particles interact with the nodes of a grid. Also, cache locality can be violated over time because of particle movement. These problems can be solved by periodical sorting of the particles. The second method is based on the idea of storing particles of the same grid cell in a separate array. This approach, due to including information about the position of each particle, addresses all the issues above, but requires special processing for the movement of particles between grid cells. Both approaches are implemented in our PIC codes PICADOR and Hi-$\chi$ (Section~\ref{subsec:performance}).

Another controversial issue concerns particles' data layouts in memory. There are two main approaches: an array of structures (AoS) and a structure of arrays (SoA). The AoS pattern provides better memory locality and is therefore more cache-friendly, while the SoA pattern is more suitable for vector processing, such as SIMD calculations on CPUs and data parallelism on GPUs. The detailed analysis of both approaches is presented in \cite{volokitin2021high}. In our codes, we employ the SoA pattern because it provides almost the same performance as AoS for scalar codes on CPUs and can show substantial speedup in vectorized codes on many CPU and GPU architectures.

The QED-PIC method may also require certain improvements to the particle ensemble data structure. Particle processing in the QED algorithm implies the relevant handling of different types of particles, therefore, it is better to store particles of different types in separate arrays to provide the opportunity for vector calculations. In the PICADOR code, particles of different types are stored together. Therefore, we need to sort them according to their type. To improve performance, we sort particles within small batches of particles (\textit{chunks}). In the Hi-$\chi$ code, we have several arrays depending on particle types, which allows us to achieve better performance.

Data structures for particles generated by QED cascades (line 3 in Algorithm~\ref{alg:run_avalanche}) deserve special attention. The classic dynamic arrays and their high-level wrappers, such as the \verb|std::vector| class from the standard C++ library, are relevant because they provide all the necessary operations with amortized constant-time complexity. However, we can adapt the vector data structure to suit our needs. For example, implementations of the standard vector class may perform many reallocations when the vector size is small, so it is profitable to allocate some extra memory at the beginning. Experiments have shown that when a QED cascade is generated, the maximum number of produced particles by a single particle is not very high and does not change significantly during the simulation. Therefore, the initial length of the vector of particles can be easily estimated. Note that in Algorithm~\ref{alg:run_avalanche} the function that returns the vector length is called frequently. Therefore, it would be beneficial to store the vector length in a separate variable rather than calculating it many times based on the starting and ending boundaries of the vector. Such an approach is implemented in Intel Compiler. We developed a custom vector data structure \cite{hichi} based on the standard vector from the Intel oneAPI DPC++/C++ Compiler 2023.0.0 and achieved a 30\% speedup of the QED module. 

\subsection{Multi-threading}\label{subsec:multithreading}


Parallel processing of an ensemble of particles in the QED-PIC method requires solving many performance problems. Firstly, we need to take into account the particle ensemble data structure and employ load balancing methods on distributed and shared memory {\cite{germaschewski2016plasma, miller2021dynamic, fedeli2022picsar, surmin2015dynamic, meyerov2019exploiting}}. Secondly, the QED method requires some specific solutions related to generating new particles, which must be saved in the particle ensemble appropriately. Besides, if a particle array is processed in parallel, data conflicts may occur when different threads add particles to the array simultaneously. We solve these problems by saving newly generated particles to separate arrays for each thread. It causes some overhead when we merge these arrays into the particle ensemble, but gives us additional opportunities. For example, before merging arrays, we can filter out the photons with low energy, that are unable to produce pairs, or perform thinning and merging procedures on particles.


\subsection{Vectorization}\label{subsec:vectorization}

Vectorized computations on modern architectures provide excellent performance opportunities. C, C++ and Fortran compilers become more and more sophisticated in auto-vectorization, but there are still many cases where they cannot vectorize a code. Naive implementation of Algorithm~\ref{alg:particle_push} contains many dependencies that prevent autovectorization. Improved implementation more suitable for vectorization is presented in Algorithm~\ref{alg:vec_particle_push}.

Firstly, vectorization requires avoiding function calls within the vectorized loop. Most of the functions that are called in the \verb|ParticlePush| procedure can be successfully inlined, so they do not prevent  vectorization. For example, the function that calculates the event rate using the polynomial approximation can be inlined and then vectorized because it performs a certain fixed number of arithmetic operations. Mathematical functions, such as \verb|log|, can be replaced with their vector analogues by a compiler. However, some functions, like the \verb|random()| function that calls the random generation procedure, can not be performed in a vectorized form. In order to solve this problem, we split the loop into several smaller loops and call these functions in scalar loops before or after the main calculations. The temporary results are stored in local static arrays with the fixed size of a chunk.

Secondly, the \verb|ParticlePush| procedure includes a \verb|while| loop, which cannot be vectorized because of an unknown number of iterations combined with the if-statement determining the end of the loop. Statistics from the test case (see Section~\ref{subsec:problem}) showed that, for 95\% of the particles, the \verb|while| loop performed only one iteration. Therefore, we vectorize the code on the particle array level. The \verb|ParticlePushChunk| procedure (Algorithm~\ref{alg:vec_particle_push}) processes a \textit{chunk} (batch) of particles. The chunk size should be a compile-time constant, so that a particle chunk can be cached. It is beneficial to choose a chunk size of 32, 64 or 128 particles. We change the order of operations and perform all particle calculations within the \verb|while| loop for each chunk. After verifying the exit condition of the loop, we rearrange particles in the chunk so that the particles that are finally processed are located at the beginning of the array and the particles that require more iterations are placed at the end. This sorting procedure can be done in linear time. In the next iteration of the \verb|while| loop, only the second part of the array is processed.

\begin{algorithm}[!ht]
\caption{The vectorized particle push algorithm for a chunk of
charged particles with account for QED effects}\label{alg:vec_particle_push}
\lstset{language=pascal,texcl=true,numbers=left,basicstyle=\small\sf,commentstyle=\small\rm,mathescape=true}
\begin{lstlisting}          
procedure ParticlePushChunk($particleChunk$, $chunkSize$, \
        $generatedPhotons$, $t_{final}$)
    while $chunkSize$ > 0 do
        for i = 0 to $chunkSize$ do
            $randomNumberChunk$[i] = random()
#pragma simd
        for i = 0 to $chunkSize$ do
            $\chi$[i] = Calculate$\chi$($particleChunk$[i])
            $\lambda$[i] = ComptonRate($\chi$[i])
            $\Delta \hat t$[i] = -log($randomNumberChunk$[i])/$\lambda$[i]
            if $particleChunk$[i].time + $\Delta \hat t$[i] > $t_{final}$ then
                $\Delta \hat t$[i] = $t_{final}$ - $particleChunk$[i].time 
            BorisPusher($particleChunk$[i], $\Delta \hat t$[i])
            $particleChunk$[i].time += $\Delta \hat t$[i]
        $nHandled$ = SortHandledParticles($particleChunk$)
        $particleChunk$ = $particleChunk$[$nHandled$:$chunkSize$]
        $chunkSize$ -= $nHandled$
#pragma simd
        for i = 0 to $chunkSize$ do
            $newPhotonChunk$[i] = PhotonEmission($particleChunk$[i])
        $generatedPhotons$.Add($newPhotonChunk$[0:$chunkSize$])
\end{lstlisting}
\end{algorithm}

\section{Experiments}\label{sub:experiments}

\subsection{Test Problem}\label{subsec:problem}

In order to test procedures optimized for simulations of QED processes we consider a fundamental problem of QED cascade development in extremely strong laser fields \cite{bell.prl.2008} which can be attained at forthcoming laser facilities \cite{danson.hplse.2019}. A QED cascade is a seeded field-assisted avalanche-like process based on gamma-photon generation by accelerated electrons/positrons and on the decay of gamma-photons into electron-positron pairs. The complexity of analysis of the cascade lies in the stochastic nature of QED processes and intricate functional dependencies of process rates on particle momentum and on fields at the particle position \cite{nikishov.spjetp.1964}. An additional complicating factor is the necessity of taking into account particle motion and field evolution which can significantly influence the spatio-temporal evolution of QED processes. This factor increases the sensitivity of the QED cascade simulation to the accuracy of the QED-PIC method.

The chosen field configuration for our tests is the plane standing circularly polarized wave. The QED cascade is well studied in this basic field configuration \cite{bell.prl.2008,fedotov.prl.2010,bashmakov.pop.2014,grismayer.pre.2017,bashinov.pra.2017}. On the one hand, it enables the spatial pair plasma distribution to be a criterion of the validity of simulations, in addition to the cascade growth rate. On the other hand, the field configuration allows using periodic boundary conditions which sufficiently reduce computational costs and enable efficient one-node testing.

Without loss of generality let us assume that the electric field of the standing wave rotates in the YZ plane. We choose the wave amplitude equal to $6.25\times10^{15}$~V~m$^{-1}$. At this amplitude both particle generation and particle motion are significant to the cascade \cite{bashinov.pra.2017} and the spatial distribution of pairs is a measure of the accuracy. The wavelength $\lambda$ is 800~nm, which agrees with the one planned at forthcoming laser facilities \cite{danson.hplse.2019}. The corresponding frequency $\omega_0$ is $2.35\times10^{15}$~s$^{-1}$ and the wave period $T$ is $2.67\times10^{-15}$~s.

The simulation box is a cube with an edge of $1\lambda$. The number of cells is 128 along the X axis and 4 along transverse axes. Note that the wave is plane and fields are homogeneous in directions perpendicular to the X axis. The electric field antinode is located at the cube center. The time step is $0.005T$.

Initially uniformly distributed seed electrons and positrons are at rest and fill the simulation box. Since for the test we focus only on the well-studied linear stage of the QED cascade (fields can be considered given), the initial particle density $n_0$ influences only the duration of the linear stage. For the chosen $n_0=10^{9}$~cm$^{-3}$ the duration of the linear stage is around $20T$. The growth of particle quantity becomes exponential with high accuracy after around $3.5T$. We employ the resampling method "global leveling" \cite{muraviev.cpc.2021} to avoid memory shortage and to maintain high performance. Cascade development is simulated within $t_\mathrm{sim}=5T$ and for performance analysis within $t_\mathrm{sim}=5T$ and $20T$. In order to reduce the variance of results we choose a quite large initial quantity of macro particles, $2.5\times10^5$.

\subsection{Validation}\label{subsec:validation}

\begin{figure}[h]
\centering
\includegraphics[width=1.0\textwidth]{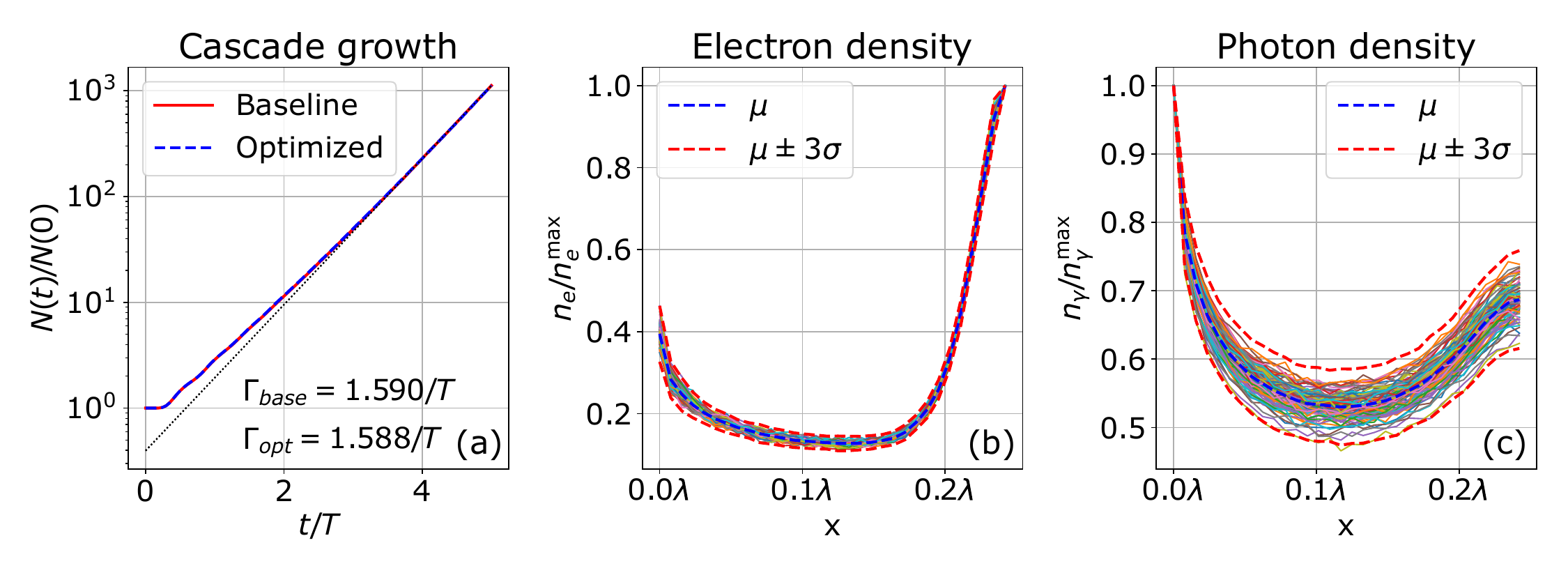}
\caption{(a) Cascade growth for the test problem of cascade development. Results are shown for two QED implementations: baseline and optimized. (b,c) Particle density is given at $t=5T$. We only demonstrate a quarter wavelength along the $x$-axis due to the symmetry of the distribution. Colored solid lines show electron and photon densities for a series of 100 simulations. Positron density is similar to electron density, so it is not displayed here. Dashed lines show the mean density $\mu(x)$ and the maximal deviation according to the empirical rule $3\sigma$.}\label{fig:verification}
\end{figure}

In this section, we describe the technique we used to compare the results of two QED simulations and justify the correctness of our optimized QED implementation. The specificity is that two QED simulations produce qualitatively similar but not exactly matching results due to randomness in the physical nature of the events, and in the algorithm. In Figure~\ref{fig:verification}(b,c), we demonstrate the possible particle distributions at a certain point in time during cascade development in the test problem described in Section~\ref{subsec:problem}. We conducted a series of 100 QED simulations and estimated the sample mean and standard deviation, assuming that the results follow a normal distribution. According to the results, the relative deviation from the mean photon density is 5-10\%, and it increases with the number of iterations. The same effect occurs when two different implementations of the QED algorithm, differing only in the order of operations, are compared with a fixed seed.

In order to compare simulations performed using the reference implementation of the QED module and the optimized one, we consider the cascade growth parameter $\Gamma$. The number of electrons $N(t)$ in a QED simulation increases as $\sim e^{\Gamma t}$, where $\Gamma$ characterizes the rate of cascade development and may be used to compare the results of two QED simulations. The experiments show that the relative discrepancy of growth rates of the reference QED implementation and the optimized QED implementation $\eta_\Gamma=(\Gamma_{base}-\Gamma_{opt})/\Gamma_{base}$ is less than 0.5\% (Figure~\ref{fig:verification}a).

\subsection{Performance}\label{subsec:performance}

The experiments described in this section were performed on a supercomputer node with 2x Intel Xeon Silver 4310T CPU (10 cores, hyper-threading, 2.3 GHz, AVX512). The code was built with the Intel oneAPI DPC++/C++ Compiler 2023.0.0. We implemented the QED module in two our frameworks: PICADOR and Hi-$\chi$. PICADOR is a fully parallel 3D QED-PIC code capable of running on heterogeneous cluster systems. The code has a flexible modular structure and can be extended to simulate various physical effects. The High-Intensity Collisions and Interactions (Hi-$\chi$) project \cite{hichi} has been developed over the past few years. It is designed based on the experience of developing the PICADOR code and aims to create an environment for performing simulations and data analysis in the research area of strong-field particle and plasma physics. Hi-$\chi$ combines the benefits of rapid prototyping in Python with the high performance of key kernels in C++. Hi-$\chi$'s flexible architecture allows us to quickly test different performance optimization approaches, reducing code development time.

Initially, we carried out the series of optimizations presented in Sections \ref{subsec:data_structures}-\ref{subsec:vectorization} on the Hi-$\chi$ code. Firstly, we developed a baseline multithreaded version of the QED module, taking into account the aspects discussed in Section \ref{subsec:multithreading}. The multithreaded code provided a 3x speedup compared to single-threaded mode  on the test problem described in Section \ref{subsec:problem} and 100 iterations of the particle-in-cell method. Given the complexity of the code, this result is considered satisfactory. This version of the QED module accounted for about 80\% of the total simulation time (Table \ref{tb:hichi_preformance}). Secondly, we applied the optimizations concerning data structures and vectorization. The optimized container for newly generated particles (Section \ref{subsec:data_structures}) provided 30\% speedup in the execution time of the QED module. Vectorization (Section \ref{subsec:vectorization}) gave an additional 2x speedup. Overall, we achieved a 2.4x speedup (61\% gain) compared to the parallel QED version, and the total simulation time improved by 2.2x (56\% gain).

\begin{center}
\begin{table}[h]
\caption{Time of different optimization stages of the QED module in the Hi-$\chi$ framework}\label{tb:hichi_preformance}
\begin{tabular}{ p{2.7cm}||p{1.8cm} p{1.8cm}|p{1.8cm} p{1.8cm} }
 \hline
 \multirow{2}{*}{Code version} & \multicolumn{2}{c|}{Time, s} & \multicolumn{2}{c}{Speedup} \\
 \cline{2-5}
 & QED only & Total\newline simulation & QED only & Total\newline simulation \\
 \hline
 \hline
 Sequential mode & 18.74 & 23.44 & - & - \\
 Multithreading & 6.43 & 7.69 & 3.0x & 3.0x \\
 Data structures & 4.42 & 5.71 & 1.4x & 1.3x \\
 Vectorization & 2.19 & 3.48 & 2.0x & 1.6x \\
 \hline
 \textbf{Total gain} & - & - & \textbf{8.5x} & \textbf{6.7x} \\
 \hline
\end{tabular}
\end{table}  
\end{center}

The developed optimizations for the Hi-$\chi$ code were also ported to the PICADOR code. Since that PICADOR is intended for parallel simulations of large-scale physical problems on supercomputers, we increased the simulation duration from 100 iterations of the PIC method to 400 iterations and compared the performance of the multithreaded QED version. Compared to the baseline parallel version of the QED method, all the particle modules including the optimized QED showed a 30\% acceleration, and the total simulation time was 25\% shorter. While the gain is not as significant as for the Hi-$\chi$ code, due to differences in the code design and particle ensemble data structures, it is still substantial improvement for our purposes.

\section{Conclusion}\label{sec:conclusion}

The paper addresses the problem of inefficient use of computing resources in the numerical simulation of QED cascades with the previously presented method \cite{volokitin2023optimized}, even though it eliminates unnecessary time step sub-cycling. We found that appropriate changes to data structures storing the particle ensemble, and the improved algorithm better suitable for vectorization and shared-memory parallelization, enhance performance by tens of percent. Performance testing was carried out on only one cluster node, however, scaling the simulation to several nodes does not directly interfere with the optimizations performed, and therefore does not affect the interpretation of the results. Our improvements are integrated into two codes, namely PICADOR and Hi-$\chi$, and demonstrate the robustness of the result. We hope that our methods can be used in other PIC codes. The QED module is publicly available \cite{hichi}.

\backmatter


\section*{Acknowledgements}
The authors acknowledge the use of computational resources provided by the Lobachevsky University and Joint Supercomputer Center of RAS. We acknowledge the support of the Ministry of Science and Higher Education of the Russian Federation, projects FSWR-2023-0034 (VV and IM) and FFUF-2024-0030 (AB and AM).




\setlength{\bibsep}{0.13cm}
\bibliography{sn-bibliography}

\end{document}